\documentclass[twocolumn,aps,english,prb,showpacs,floatfix]{revtex4-1}

\usepackage[colorlinks=true,urlcolor=blue,citecolor=blue,linkcolor=blue]{hyperref}

\usepackage{amssymb}
\usepackage{graphicx}
\usepackage{amsmath,color}
\usepackage{color}

\renewcommand{\>}{\rangle}
\renewcommand{\(}{\left(}
\renewcommand{\)}{\right)}
\renewcommand{\[}{\left[}
\renewcommand{\]}{\right]}

\DeclareMathOperator{\Tr}{Tr}
\newcommand{\Id}{\mathbb{I}}

\newcommand{\eqnref}[1]{(\ref{#1})}

\newcommand{\Sd}{S^\mathrm{d}}
\newcommand{\Sc}{S^\mathrm{c}}
\newcommand{\Sb}{S^\mathrm{b}}

\usepackage[sort&compress]{natbib}
\setcitestyle{numbers,square}

\makeatletter

\begin{document}

\title{Many-body localization in the presence of a small bath}

\author{Katharine Hyatt}
\author{James R. Garrison}
\affiliation{Department of Physics, University of California, Santa Barbara, California 93106, USA}

\author{Andrew C. Potter}
\affiliation{Department of Physics, University of California, Berkeley, California 94720, USA}

\author{Bela Bauer}
\affiliation{Station Q, Microsoft Research, Santa Barbara, California 93106, USA}

\date{\today}

\begin{abstract}
In the presence of strong disorder and weak interactions, closed quantum systems can enter a many-body
localized phase where the system does not conduct, does not equilibrate even for arbitrarily long times,
and robustly violates quantum statistical mechanics.
The starting point for such a many-body localized phase is usually taken to be an Anderson insulator where, in the limit of vanishing
interactions, all degrees of freedom of the system are localized. Here, we instead consider a model where
in the non-interacting limit, some degrees of freedom are localized while others remain delocalized.
Such a system can be viewed as a model for a many-body localized system brought into contact with a
small bath of a comparable number of degrees of freedom.
We numerically and analytically study the effect of interactions on this system and find that generically, the entire system delocalizes. However, we
find certain parameter regimes where results are consistent with localization of the entire system, an effect recently
termed many-body proximity effect.
\end{abstract}

\maketitle

\section{Introduction}

While it has become clear in recent years that the eigenstate thermalization hypothesis (ETH)~\cite{Deutsch91,Srednicki94} provides
the correct picture for the emergence of quantum statistical mechanics in broad classes of closed quantum systems~\cite{Rigol08},
the phenomenon of many-body localization (MBL)~\cite{Anderson58,Basko06a,Basko06b,gornyi2005} has appeared as a scenario
where quantum statistical mechanics is robustly violated.
By now, overwhelming numerical~\cite{Oganesyan07,Znidaric08,Pal2010,Bardarson12,Iyer12,bauer2013,kjall2014,luitz2015}, analytical~\cite{imbrie2014}
and experimental~\cite{DeMarco13,Bloch15,smith2015,bordia2015}
evidence has been amassed that a many-body localized phase exists in strong disorder and for finite-strength interactions
(for a recent review, see Ref.~\onlinecite{nandkishore2015-review}).
The violation of ETH
most prominently manifests itself in an area law for the entanglement entropy in highly excited eigenstates~\cite{bauer2013}: unique to an MBL system,
the entanglement entropy of a region scales only with the size of the boundary of that region in almost all states of
the many-body energy spectrum. Other key properties of MBL phases include a discrete local spectrum and vanishing conductivity~\cite{Basko06a,Basko06b},
a logarithmic growth of entanglement entropy~\cite{Znidaric08,Bardarson12,Serbyn2013-1}, and a complete set of local integrals of motion that describe the entire
many-body spectrum~\cite{Huse2013,Serbyn2013}.

The description of many-body localization generally assumes a system where, in the limit of vanishing electron-electron interactions,
the system becomes an Anderson insulator in which all single-particle states are localized. The eigenstates of a many-body
localized system are connected to the eigenstates of the Anderson insulator by a finite-depth local unitary transformation~\cite{bauer2013}.
Recent work~\cite{li2015,modak2015} has raised the question of whether MBL can also exist in a system where
in the non-interacting limit, a critical single-particle energy, the mobility edge, separates localized and delocalized states.
In one dimension, this can be achieved in certain types of quasi-disordered systems, and it is a generic scenario in higher dimensions.
For a particular incommensurate potential, Ref.~\onlinecite{li2015} found three regimes: a many-body localized phase; an ergodic phase;
and an intermediate phase that exhibits volume-law entanglement scaling but violates eigenstate thermalization in a more subtle way
by having a large eigenstate-to-eigenstate variance of the expectation value of local operators, even in narrow energy windows.
These regimes are separated by many-body mobility edges, i.e.\ critical energy densities separating localized from delocalized states in the
many-body spectrum. The existence of such many-body mobility edges has been suggested based both
on analytical~\cite{Basko06a,Basko06b} and numerical~\cite{kjall2014,luitz2015} observations; however, recent
work has raised concerns about the stability of such a scenario~\cite{deroeck2015}.

A closely related question concerns the stability of many-body localization in open quantum systems, i.e.\ when coupled to bath degrees
of freedom. Usually a bath is taken to be very large and the backaction of the system onto the bath is neglected. In this case (as well as
in the presence of dissipation~\cite{levi2015,fischer2015}), one
expects that the effects of many-body localization will be destroyed, although there is evidence for a crossover into a regime where some
signatures of localization persist~\cite{Nandkishore2014,Gopalakrishnan2014,Johri2015}.
However, one can also consider the case where the number of degrees of freedom in the system and the bath is comparable
and the backaction may be important. In this case, the interesting possibility arises that the backaction of the MBL system may be strong enough to induce localization in the bath~\cite{nandkishore2015}. The results of Ref.~\onlinecite{li2015} may be interpreted as evidence for such a scenario.

\begin{figure}
  \centering
  \includegraphics[width=0.6\columnwidth]{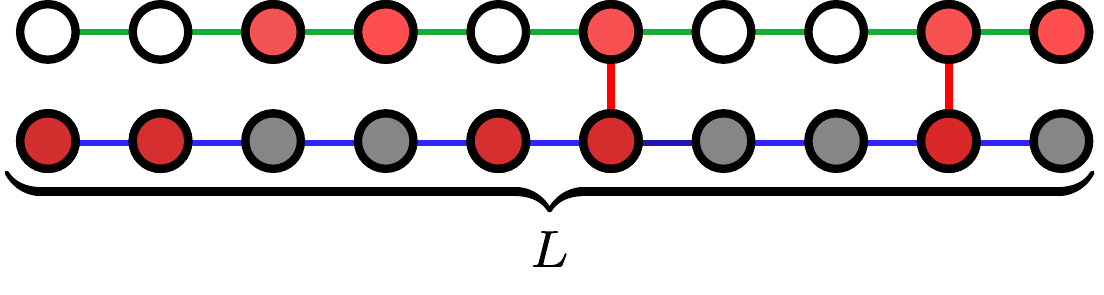}
  \caption{Schematic illustration of our setup. Here, the disorder potential only acts on the lower chain of the ladder ($\alpha=0$),
  whereas the fermions in the upper chain are affected by the disorder only through the interactions. The fermions, indicated as
  red dots, hop along each chain, and interact only through a repulsive density-density interaction on each rung.
  \label{fig:ladder} }
\end{figure}

In this paper, we investigate these questions by numerically and analytically studying a model for spinless fermions on a ladder, as sketched in Fig.~\ref{fig:ladder}.
We introduce an uncorrelated
disorder potential on one chain of the ladder, while keeping the other chain translationally invariant, and forbid hopping between the chains such that
in the non-interacting limit, the chains are completely decoupled. We then introduce a local density-density interaction on each rung.
With interactions, two sharply distinct scenarios appear: in one, the energy transport through the clean chain is sufficient
to trigger delocalization of the entire system. In the other scenario, the localized fermions -- through the density-density interaction -- act
as effective disorder potential for the fermions in the clean chain, inducing their localization.
To distinguish these two scenarios, we will consider the entanglement entropy of highly excited eigenstates as well as the time evolution
of the entanglement entropy. We find that in our model, both scenarios can be realized depending on the parameters of the system.
We will comment on other possible intermediate scenarios at the end.

Our model is closely related to the model of Ref.~\onlinecite{yao2014}, where the disorder potential is absent but the hopping strength in the
two chains is vastly different and interactions between the chains are very strong.
This model was studied in the context of dynamical effects akin to many-body localization in systems without
explicit translational symmetry breaking in the Hamiltonian~\cite{schiulaz2014,hickey2014,deroeck2014} (see also~\cite{grover2014-qdl}),
but where the initial conditions break translational symmetry.
It was observed that the time evolution starting from random product state configurations exhibits slow dynamics at an intermediate time scale,
yet relaxes at the longest time scales, consistent with with the formation of a ``quasi many-body localized" (qMBL) regime.
Furthermore, the system shows a diverging susceptibility towards spin glass ordering upon introducing disorder.
Our results are complementary in that we consider the case of strong disorder and weak interactions, and focus primarily on eigenstate rather than dynamical
properties. We discuss the relationship between the models in more detail towards the end of Section~\ref{sec:numerics}.

The remainder of this paper is structured as follows: In Sec.~\ref{sec:model}, we describe the model, our diagnostics and the numerical
approach in more detail. In Sec.~\ref{sec:pert}, we discuss a perturbative analysis of the interchain coupling. In Sec.~\ref{sec:numerics},
we describe our numerical results, and conclude in Sec.~\ref{sec:concl}.

\section{Model and numerical approach}
\label{sec:model}

The Hamiltonian for the system is (see Fig.~\ref{fig:ladder})
\begin{align} \label{eqn:H}
    \hat{H} =& -\sum_\alpha t_\alpha \sum_{i = 1}^L \left( \hat{c}^\dag_{\alpha, i} \hat{c}_{\alpha, i+1} + \text{h.c.} \right) \\ \notag
&+ \sum_{i = 1}^{L} w_i \hat{n}_{d,i} + V\sum_{i = 1}^{L}\hat{n}_{d,i} \hat{n}_{c,i}.
\end{align}
Here, $c_{\alpha,i}^\dagger$ creates a fermion on the upper, clean ($\alpha=c$) or lower, disordered ($\alpha=d$) layer. The local potential
$w_i$, acting only on the $\alpha=d$ fermions in the lower layer, is drawn uniformly from the range $[-W,W]$.
$V$ is a density-density interaction between the two chains.
Each chain has length $L$ such that the total number of sites in $2L$.
Note that the particle number on each chain is separately conserved, reducing the size of the
many-body Hilbert space. For even $L$, we choose each chain to be half-filled.

While the model is phrased here in terms of a ladder, it is equivalent to a system of two different flavors of fermions where
only one flavor experiences the disorder potential. Such a description may be applicable to experiments on cold atoms which
use different types of atom (non-convertible fermions), or different states of the same atom.
We also note that since hopping between the chains is forbidden,
the model can be mapped to a local model of hard-core bosons or spins by means of a Jordan-Wigner transformation. Since it is
also possible to apply the Jordan-Wigner transformation to only one chain of the ladder, the model is related
to spin and charge degrees of freedom in a Hubbard chain.

In Eq.~\eqnref{eqn:H}, we have not included interactions between fermions on the same chain. We have confirmed numerically
that adding a repulsive nearest-neighbor interaction between fermions on the order of the interlayer coupling or weaker 
on the same chain does not qualitatively affect the
results. We have also verified that making the strength of the inter-chain interaction random does not affect the results.

We solve for highly excited eigenstates of Eq.~\eqnref{eqn:H} in the middle of the many-body spectrum using the
shift-and-invert method, which solves for low-lying states of
\begin{equation}
    \hat{A}=\left( \hat{H} - \lambda \Id \right)^{-1}
\end{equation}
where $\lambda$ is a target energy. This approach was first used in the context of many-body localization in Ref.~\onlinecite{luitz2015}.
We use the implementation of  SLEPc~\cite{Hernandez:2005:SSF} \& PETSc~\cite{petsc-user-ref,petsc-efficient,MUMPS01,MUMPS02}, and rely on its direct LU solver
and MUMPS.
The LU factorization is used as a direct solver to perform the inversion after shifting.
Once the inverse has been computed, the Lanczos method~\cite{lanczos1950iteration} can be used to target low-lying
states of shift-and-inverted matrix. These states are the ones closest in energy to the
target $\lambda$.
We average over 250 eigenstates for 500 disorder realizations each for system sizes $L=6$ and $L=8$.
For $L = 10$, we only compute eigenstates for 50 disorder realizations.
We choose the target energy $\lambda = VL/4$, which is close to the center of the many-body spectrum. For small
systems, we have verified through a full diagonalization of $\hat{H}$ that the states thus obtained are representative
of the ``infinite-temperature'' ensemble.

\subsection{Eigenstate entanglement}

The eigenstate thermalization hypothesis~\cite{Deutsch91,Srednicki94,Rigol08} suggests that in an eigenstate of a
generic quantum system $H$ at a finite energy density $\epsilon$ above the ground state, the reduced density matrix
$\rho_\mathcal{A}$ for some region $\mathcal{A}$ will be close to a Gibbs state of the same Hamiltonian at some inverse
temperature $\beta(\epsilon)$, $\rho_\mathcal{A} \approx \exp (-\beta H_\mathcal{A})$~\cite{garrison2015}.
Among many other things, this implies that
the entropy $S(\rho_\mathcal{A}) = - \Tr \rho_\mathcal{A} \log \rho_\mathcal{A}$ will exhibit a volume law,
$S(\rho_\mathcal{A}) = s_\text{th}(\epsilon) \mathrm{vol}(\mathcal{A})$. In the center of the many-body energy band,
where $\beta \rightarrow 0$, one expects that the entropy density is close to its maximal value as given by the density
of degrees of freedom.

\begin{figure}
  \centering
  (a) \\
  \begin{tabular}{c|c|c}
  \includegraphics[width=0.3\linewidth]{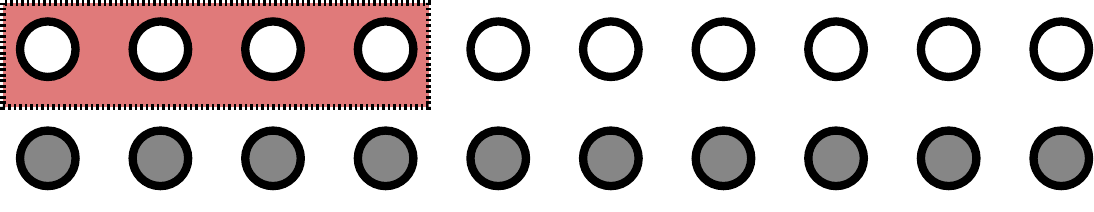}
  & \includegraphics[width=0.3\linewidth]{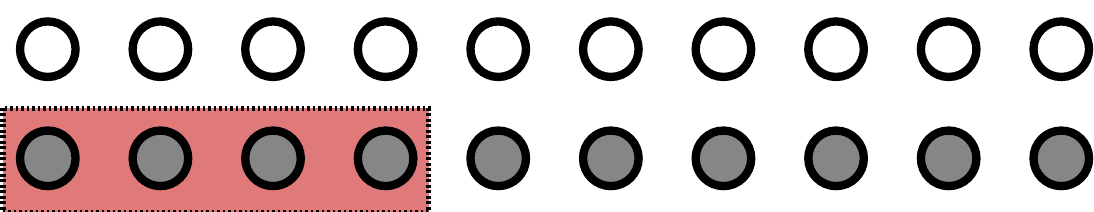}
  & \includegraphics[width=0.3\linewidth]{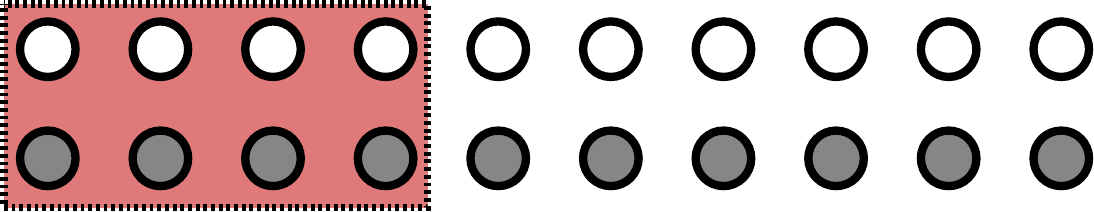} \\
  $\Sd$ &$\Sc$ &$\Sb$
  \end{tabular}
  \vspace{0.3in} \\

  (b) \\
  \includegraphics[width=\columnwidth]{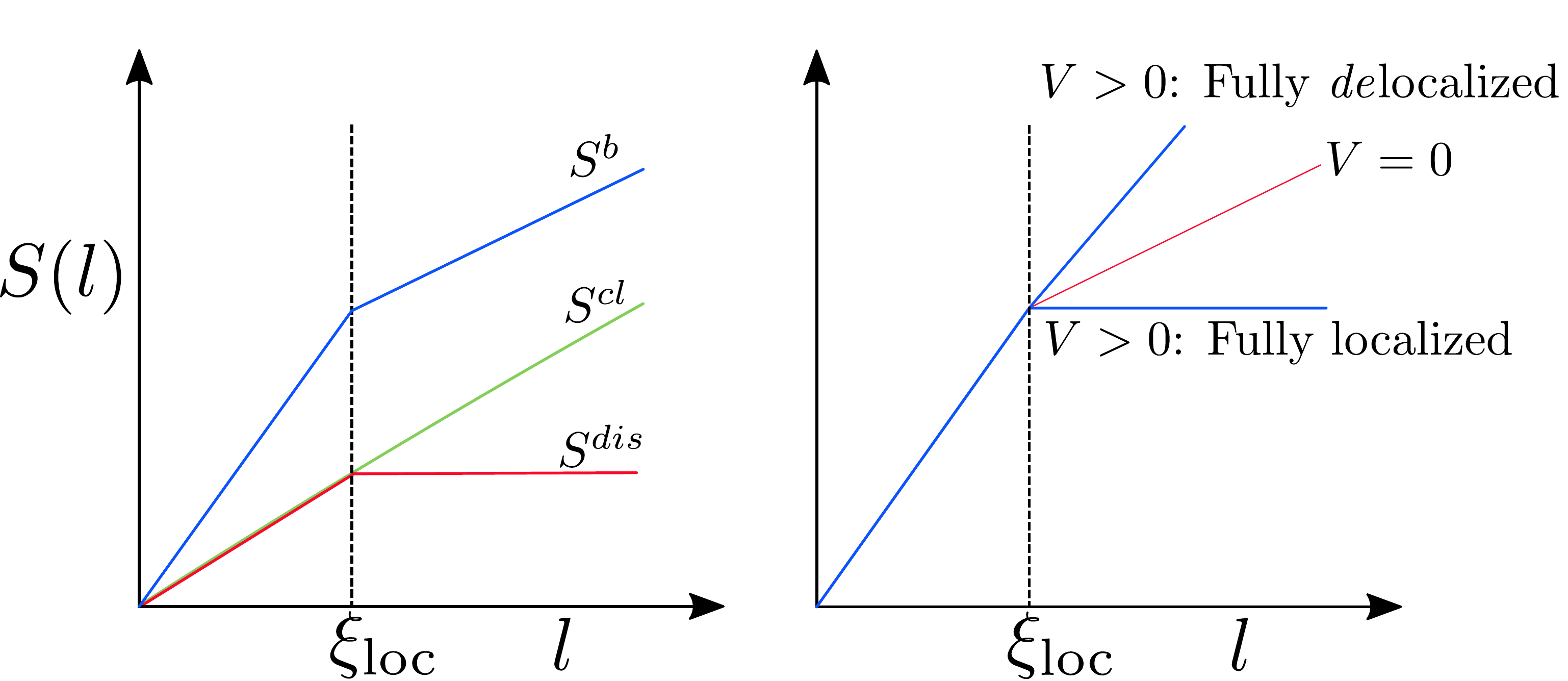}
  \caption{(a) Different entanglement cuts considered in this paper. Here, the empty circles denote sites
  in the disordered chain, while filled circles denote sites in the clean chain. From left to right, we denote
  these entropy cuts as $\Sd$, $\Sc$, $\Sb$.
  (b) Schematic behavior of the entanglement entropy. Left panel: decoupled case ($V=0$), where $\Sb=\Sd+\Sc$.
  Right panel: Interacting case ($V \neq 0$).
  \label{fig:entropyillustration} }
\end{figure}

One of the defining features of many-body localization is that this volume-law scaling is robustly violated~\cite{bauer2013}.
Indeed, the excited eigenstates of an MBL system exhibit an area law~\cite{Eisert2010}:
the bipartite entanglement entropy between some region $\mathcal{A}$ and the rest of the system
is found to scale only with the area of the boundary separating the regions. In $d$ dimensions, this leads to the scaling $S \sim L^{d-1}$,
as opposed to the volume law $S \sim L^d$ that is expected for generic systems that obey eigenstate thermalization. Eigenstate
entanglement has subsequently been used as powerful criterion to identify many-body localized phases~\cite{bauer2013,kjall2014,luitz2015}.

We use eigenstate entanglement as diagnostic of whether the system described by Eq.~\eqnref{eqn:H} delocalizes or localizes.
On a coarse level, distinguishing between a volume law and an area law allows us to determine whether the system is localized
or not. Beyond this, if we find that the system exhibits a volume law, we can consider the entropy density to determine whether
all microscopic degrees of freedom participate or whether some degrees of freedom remain localized.

Specifically, we consider the entanglement cuts illustrated in the top panels of Fig.~\ref{fig:entropyillustration}. The cuts correspond to a contiguous block
of sites in the clean (disordered) chain, which we label as $\Sc$ ($\Sd$), as well as a cut that contains both chains, labeled as $\Sb$.
In the generic case where the two chains are coupled, an area law will only appear in $\Sb$, since for the other cuts the area of the
boundary scales with the volume of the block. However, in the non-interacting case $V=0$ where the two chains are completely decoupled,
the entropy of each of the chains separately can provide valuable insights.

The bottom panels of Fig.~\ref{fig:entropyillustration} schematically illustrate the behavior of the various entropy cuts.
The lower left panel shows the case of decoupled, non-interacting chains ($V=0$): the delocalized fermions in the clean layer contribute a volume law, $\Sc = s_\text{th} L$
where $s_\text{th} \approx \log 2$ is the thermal entropy density of a single layer at infinite temperature. The localized layer, on the
other hand, exhibits a volume law only for blocks smaller than the localization length $\xi_\mathrm{loc}$~\cite{grover2014} and saturates to a
constant beyond that. The total entropy, $\Sb=\Sd+\Sc$, thus shows a volume law with prefactor $2 s_\text{th}$ for
$l < \xi_\mathrm{loc}$, and crosses over to a volume law with a reduced prefactor $s_\text{th}$ for $l > \xi_\mathrm{loc}$.

In the interacting case, the two scenarios discussed above are localization and delocalization of the entire system. These are illustrated
in the lower right panel of Fig.~\ref{fig:entropyillustration}: If the system becomes fully localized on some scale $\xi'$, the entanglement
entropy for the joint system $\Sb$ will exhibit an area law for $l > \xi'$ (note that while the figure shows the case $\xi' = \xi_\mathrm{loc}$, this need
not necessarily be the case). If, on the other hand, the system fully delocalizes, the entropy
will show a volume law with prefactor $2 s_\text{th}$ for all scales. Thus, in either scenario a strong signature appears in the eigenstate
entanglement: the entropy either becomes constant, or the coefficient of the volume law doubles.

\section{Perturbative analysis of interchain coupling}
\label{sec:pert}

We expect that the eventual fate of the localization or thermalization of the ladder rests on the outcome of a competition between the tendency for the disordered chain to localize the states on the clean chain and the ability of the states in the clean chain to act as a thermalizing bath for the disordered chain. Before turning to direct numerical simulations, we consider the limit of weak interactions, $V\ll t_d,t_c$, where we can work perturbatively in $V$ near the decoupled chain limit to estimate which of these effects is more important.

For $V=0$, the clean chain has no randomness and exhibits extended single-particle eigenstates. With $V\neq0$, however, if the disordered chain remains localized, the random distribution of charges $\langle \bar{n}_{d,i} \rangle$ in a given eigenstate produces an effective disorder potential $\mu_i\approx V\<n_{d,i}\>$ in the clean chain. Due to the one-dimensional nature of the system, even an infinitesimally weak random potential will produce localization. In the case of weak interactions and thus weak disorder, localization occurs via quantum interference, and we can perturbatively estimate (using Fermi's golden rule) the localization length in the clean chain to be $\xi_V \approx t_c^2 / (V^2\,\delta n_d^2)$, where $\delta n_d^2\approx \rho_d(1-\rho_d)$ and $\rho_d$ is the mean density of particles in the disordered chain.

As a technical aside, this treatment amounts to a Hartree-type approximation of the inter-chain interaction to obtain an effective disorder potential $\mu_i = V\<n_{d,i}\>$ for the clean chain. Note that Fock-type exchange self-energies are zero in this model where particle number in each chain is separately conserved.
Next, we will perturbatively incorporate interactions using the Hartree-dressed Green functions.

In the absence of this tendency towards localization, energy can propagate along the disordered chain mediated by resonant interactions with states in the clean chain. We can perturbatively estimate whether such resonant interactions can persist in spite of the tendency towards developing a finite localization length $\xi_V$ in the clean chain. The typical level spacing for particle-hole excitations in the clean chain is:
\begin{align}
\delta^{(2)}\approx \frac{1}{\Lambda_c\(\nu_c\xi_V\)^2}\approx \frac{V^4}{t_c^3}
\end{align}
where $\nu_c\approx 1/t_c$ and $\Lambda_c\approx t_c$ are the single-particle density of states and bandwidth, respectively. For strong disorder ($W\gg t_d$), the single-particle states of the disordered chain are well-localized with characteristic localization length $\xi_d\approx \[\log \(W/t_d\)\]^{-1}$. In this case, the simplest interchain interaction process is for a particle-hole excitation of two overlapping localized orbitals in the disordered chain to excite a particle-hole pair in two orbitals of the clean chain that reside within distance $\xi_V$ of those in the clean chain.

For concreteness let us label a fixed pair of orbitals in the disordered chain by $a,b$, whose wave functions have the schematic form $\phi_{a,b}\approx \frac{1}{\sqrt{\xi_d}}e^{-|x-x_{a,b}|/2\xi_d}$, and similarly denote two orbitals in the other chain by $\alpha$ and $\beta$ respectively with wave functions $\phi_{\alpha,\beta}\approx \frac{1}{\sqrt{\xi_V}}e^{-|x-x_{\alpha,\beta}|/2\xi_V}$. Then, for $|x_a-x_b|<\xi_d$, and $|x_{\alpha,\beta}-x_a|<\xi_V$ the matrix element for the interchain interaction among these orbitals is roughly:
\begin{align}
\Gamma_{(a,b);{\alpha,\beta}} &\approx V\int dx\ \phi^*_\alpha(x)\phi_\beta(x)\phi^*_a(x)\phi_b(x) \\
& \approx \frac{V}{\xi_V}\approx \frac{V^3}{t_c^2}
\end{align}
Fixing our attention on a specific pair of orbitals in the disordered chain, the number of such transitions that are resonantly connected by matrix element $\Gamma$ is of order $N_\text{res}\approx \frac{\Gamma}{\delta}\approx \frac{t_c}{V}$, which is large for weak interactions, and in fact diverges in the limit of $V\rightarrow 0$. This divergence strongly suggests that, for weak interactions, the tendency for randomness in the disordered chain to localize states in the clean chain is insufficient to prevent it from acting as a bath for the disordered chain.  The full system thus thermalizes in this limit on a timescale set by the magnitude of the matrix element $\Gamma$.

To address the fate of thermalization in this ladder beyond these perturbative considerations, we now turn to microscopically exact numerical simulations.

\section{Numerical results}
\label{sec:numerics}

\subsection{Equipotent hopping}

\begin{figure}
    \includegraphics[width=\linewidth]{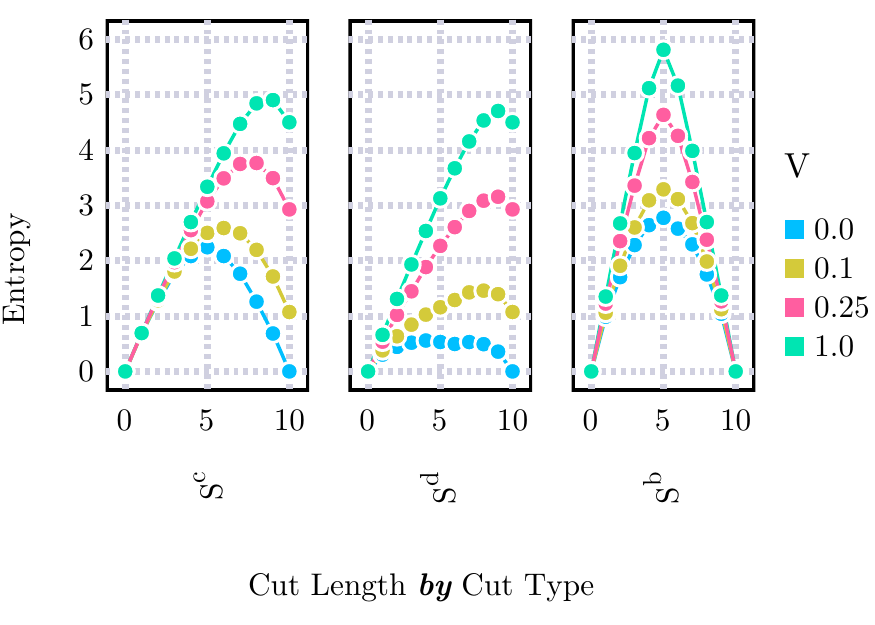}
    \caption{
    Entropy for three different cuts for a ladder of length $L=10$ with $t_0=t_c=1$ and $W=4$. The different cuts
    are illustrated in the top panels of Fig.~\ref{fig:entropyillustration}. Error bars are comparable to the marker size. }
    \label{boxpic}
\end{figure}

\begin{figure}
    \includegraphics[width=0.49\linewidth]{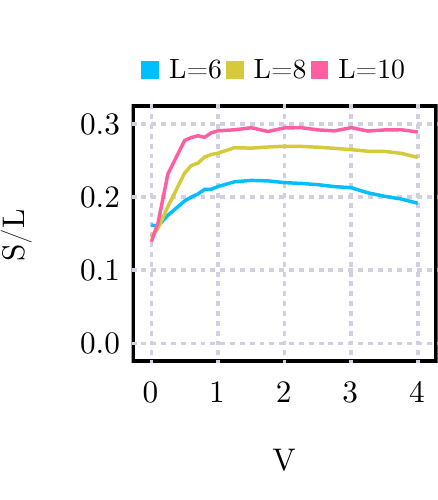}
    \includegraphics[width=0.49\linewidth]{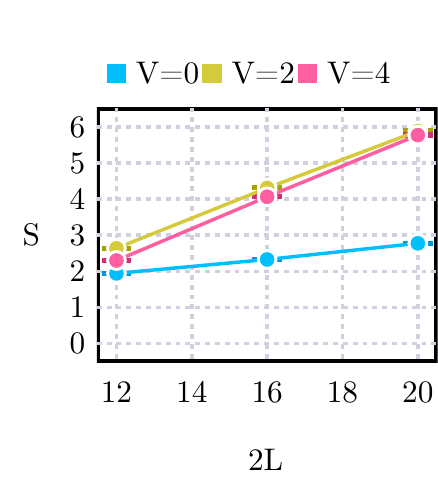}
    \caption{
    Left panel: Entropy $\Sb$ at the center of the system as a function of interlayer coupling $V$, for different
    system sizes and disorder strength $W=4$. We observe that the increase of the entropy due to adding
    weak interactions sharpens as the system size increases.
    Right panel: Entropy $\Sb(L/2)$ for different $V$ as a function of system size. The slope corresponds to
    the entropy density, which is observed to increase drastically as interactions are increased, compatible
    with all degrees of freedom contributing to the entropy.
    \label{fig:boxentropy_unit} }
\end{figure}

We first consider the case of equal hopping strength in the clean and disordered layer, $t_d = t_c = 1$. In Fig.~\ref{boxpic}, we show
the behavior of different entropy cuts in this case for a system with a total of $2L=20$ sites and disorder strength $W=4$.
First we consider the results in the non-interacting case, $V=0$. Here, we observe a volume law for $\Sc$ (left-most panel), while
the disordered system exhibits a saturation of the entanglement entropy $\Sd$ for $l \gtrsim 3$ (center panel). Note that for
$V=0$, $\Sd(l)=\Sd(L-l)$ (and similarly for $\Sc$), which is not true for $V \neq 0$.
By identifying the saturation point of the entropy for $\Sd$, we can read off that the
localization length for $W=4$ is $\xi_\mathrm{loc} \approx 3$. The right-most panel shows the joint entropy $\Sb$, which for
$V=0$ is simply the sum of the two contributions and is therefore dominated by the volume law in the clean chain.

As we turn on interactions between the two layers ($V > 0$), we observe that the entanglement entropy for each cut increases dramatically.
For the entropy cuts isolating each chain, this is expected as the boundary between the two chains begins to contribute
to the entanglement entropy. However, we note that in the limit of strong enough interactions, the entropies $\Sd$ and $\Sc$ become approximately
equal, indicating that there is no distinction between the two chains. Considering the joint entropy, we find that the entropy
at the center of the system approximately doubles from $\Sb(L/2) \approx 2.8$ to $\Sb(L/2) \approx 5.8$.
This is consistent with a volume law contribution from both chains. The maximum measured value at the center of the system is close
to the upper bound $\Sb(L/2) \approx 6.9$; the discrepancy can be attributed to finite-size corrections.
These results are strongly suggestive of delocalization of the entire system for $V > 0$.

To further investigate this, we consider the cut at the center of the system, $\Sb(L/2)$, for various system sizes and interaction
strengths as shown in Fig.~\ref{fig:boxentropy_unit}. In the left panel, we have rescaled the entropy by the system size to convert to an
entropy density. We observe that the entanglement increases rapidly as interactions are turned on for each system size.
The increase sharpens as the system size is increased, indicating that at least in this parameter regime the entire
system delocalizes for infinitesimal interaction strength $V$ in the thermodynamic limit. For small system sizes, the entropy decreases for very large
interactions, but this effect does not appear to persist to larger system sizes.
In the right panel, we analyze the finite-size dependence of the entropy at the center of the system, which clearly exhibits a volume-law
scaling. The coefficient of the volume law increases rapidly as interactions are turned on.

\subsection{Narrow-bandwidth bath}

\begin{figure}
    \includegraphics[width=\linewidth]{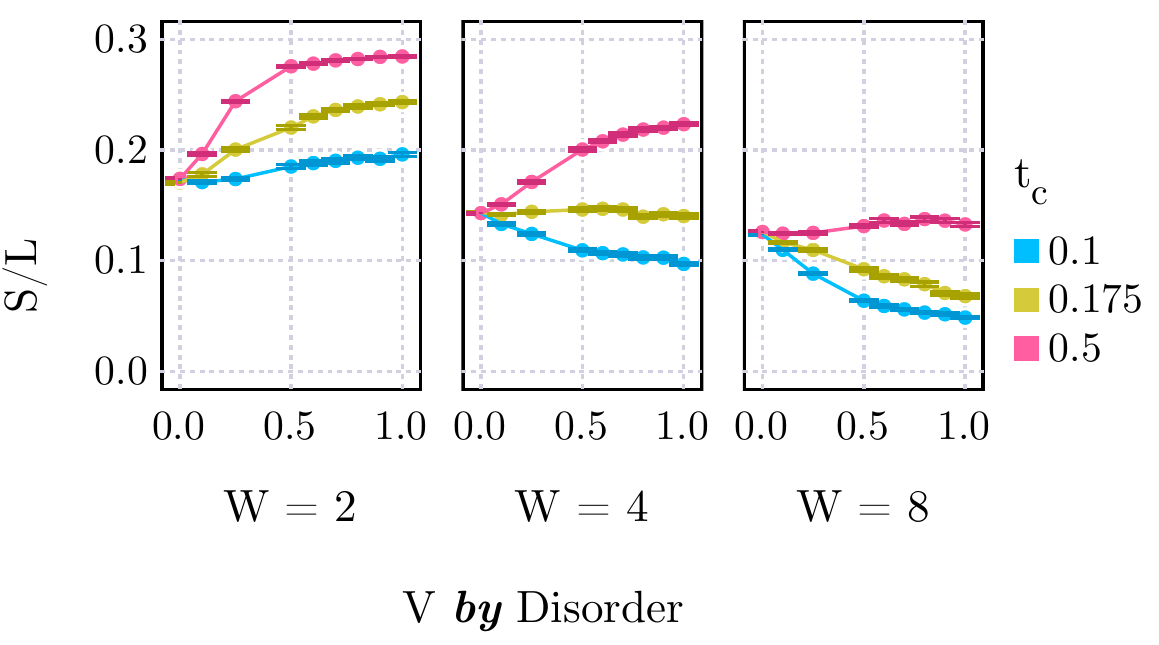}\\
    \caption{Entropy $\Sb(L/2)$ for a center cut of the system for various disorder strengths $W$ and hopping in the clean chain, $t_c$.
    As disorder is increased and the hopping $t_c$ reduced, we observe a crossover from a regime where interactions tend to
    increase entropy to a regime where interactions decrease entropy, i.e.\ drive the system towards localization.}
    \label{boxentropy}
\end{figure}

\begin{figure*}
\includegraphics[height=0.5\textheight]{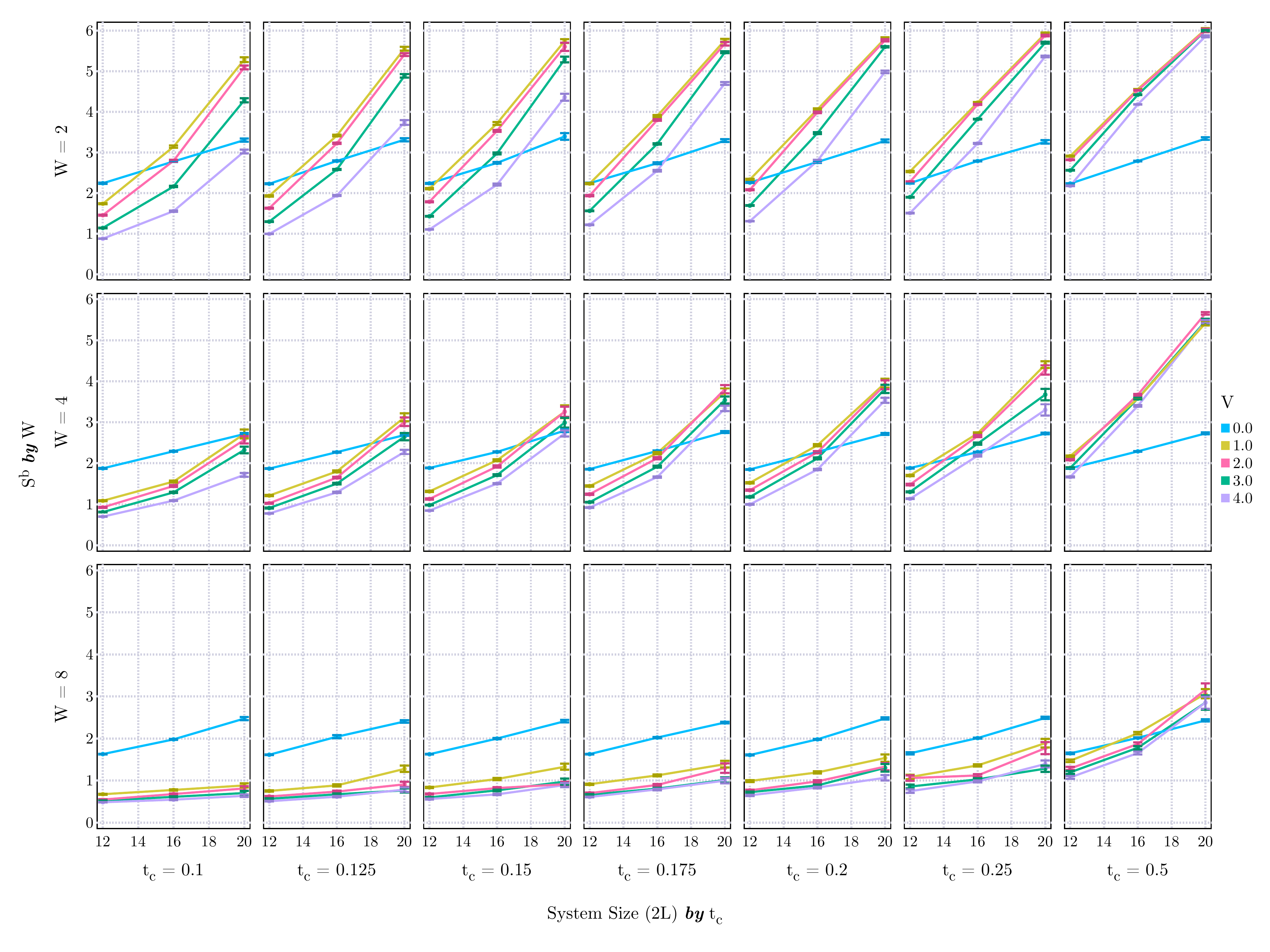}
\caption{Maximum entropy $\Sb$ for various $L$ and interlayer couplings $V$, allowing the hopping strength to vary. For very large interlayer coupling, the system appears to localize at small hopping in the clean chain.}
\label{fig:hopbox}
\end{figure*}

While the results of the previous section confirm that the system tends to delocalize, as suggested by the considerations
in Sec.~\ref{sec:pert}, it may still be possible to change the parameters of the system in such a way as to enhance
the tendency towards localization.
We focus here on the effect of tuning the hopping strength in the clean layer, $t_c$. By reducing
this hopping strength, we can reduce the bandwidth of the delocalized degrees of freedom, which reduces the amount
of energy the bath can absorb.
This regime was previously studied in Ref.~\onlinecite{Gopalakrishnan2014}.
Clearly, in a limit where $t_c \ll \delta E$, where $\delta E$ is the mean level spacing of many-body energy levels,
the system will not delocalize; however, this requires an exponentially small bandwidth $t_c \sim 2^{-L}$. In a more
physical regime, where $\delta E \ll t_c \ll t_d$ and $t_c \ll W$, one may still expect that the bath is inefficient at delocalizing
the system because its bandwidth is small compared to the energy mismatch of nearby fermion states in the disordered
system, which is of order $W$; however, higher-order resonances may invalidate this simple picture.

Our numerical results are summarized in Figs.~\ref{boxentropy} and \ref{fig:hopbox}. In Fig.~\ref{boxentropy}, we show
the entropy at the center of the system as a function of interaction strength $V$ for three different strength of the disorder and several
values of the hopping in the clean chain $t_c$, while keeping $t_d=1$. For slightly reduced values of $t_c$, such as $t_c=0.5$, we find behavior that is
very similar to the case $t_c=t_d=1$. However, for strongly reduced hopping of $t_c = 0.1$ and sufficiently strong disorder, the entropy does
not increase rapidly -- as was seen in the left panel of Fig.~\ref{fig:boxentropy_unit} -- but rather remains constant, or even decreases in the
case of strong disorder $W=8$. This implies that as interactions are turned on, the system tends towards localization rather than
delocalization.

To explore whether this is a robust effect that persists to the thermodynamic limit, we examine how this behavior depends
on the system size over the limited range available to our exact numerics. In Fig.~\ref{fig:hopbox}, we show
the entropy at the center of the system versus system sizes, for an array of hopping strengths $t_c \in [0.1,0.5]$ (keeping $t_d=1$)
and disorder strengths $W=2,4,8$.
For weak disorder ($W=2$, top row), we observe that interactions suppress the entropy for small systems and $t_c \leq 0.25$, but for sufficiently
large systems the behavior changes and the entropy of the interacting system exceeds that of the decoupled chains and the
coefficient of the volume becomes comparable to that in the case $t_c=t_d$. The scale
at which this crossover takes place depends on the choice of $W$ and $t_c$, and appears to shift to larger and larger systems
as $W$ is increased and $t_c$ reduced. For example, for $W=4$ and $t_c = 0.1$ (leftmost panel of the middle row) the crossover
scale appears to be slightly larger than the system sizes available to us.
We note that such an upturn of the entropy seems at odds with the results put forward
in Ref.~\onlinecite{grover2014}, which argued that $\partial^2 S(\ell)/\partial \ell^2 \leq 0$, where $S(\ell)$ is the entropy
of a contiguous block of $\ell$ sites in a system of total size $L$. However, this result only holds for $\ell \ll L$, whereas
here we have the case $\ell = L/2$ and therefore no clear separation between the two scales $\ell$ and $L$.

Finally, in the case of strong disorder ($W=8$), small bath bandwidth ($t_c < 0.2$), and sufficiently strong interactions,
the entropy remains strongly suppressed for all values of the interaction strength and the entire range of system
size accessible to our numerics. This is suggestive of a many-body localized phase of both chains in this limit.
However, given the finite-size crossover behavior observed away from this limit, we cannot rule out delocalization
of the system on a very long length scale.

\subsection{Time evolution}

\begin{figure}
  \includegraphics{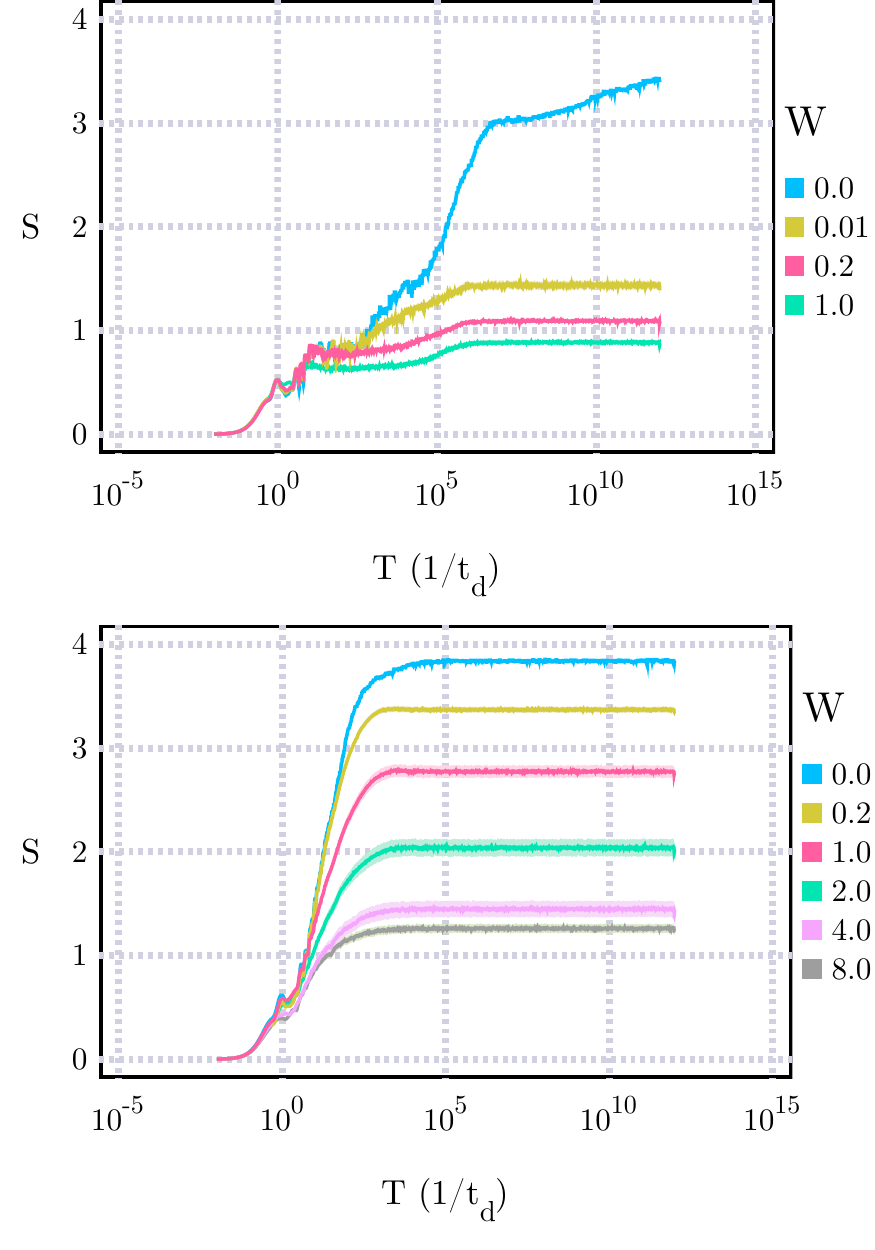}
  \caption{Time evolution of the entanglement entropy $\Sb(L/2)$ for $V=10$ and $2L=16$ total sites.
  The top panel shows data for $t_c=0.001$, corresponding to Fig. 2 of Ref.~\onlinecite{yao2014},
  while the bottom panel shows $t_c=0.1$ for comparison to Fig.~\ref{fig:hopbox}. }
  \label{fig:te}
\end{figure}

To obtain further insights into the putative localized regime at strong disorder and weak bath hopping and to connect our
results to those of Ref.~\onlinecite{yao2014}, we repeat some of the numerical experiments performed
there in the presence of weak disorder. To this end, we prepare the system in a randomly chosen product state,
evolve under the Hamiltonian~\eqnref{eqn:H} (where, following~\cite{yao2014}, we from here forward impose periodic boundary conditions),
and compute the entropy $\Sb$ at the center of the system as a function
of time $T$. We average the results over different initial states, and in the case of $W>0$ also over 10 different disorder realizations.
For each case, we smooth the data by taking the maximum over 6 adjacent time samples.
We note that these results are obtained for much stronger interaction $V$ and much smaller $t_c$ than the
results reported in the previous sections.

Our results are shown in Fig.~\ref{fig:te}. Here, the parameters for the upper panel match those of Ref.~\onlinecite{yao2014},
in particular $t_c=0.001$, $t_d=1$, and the disorder-free case $W=0$ exactly matches the data presented there.
Upon adding even a very weak disorder potential, the evolution
of the entropy for large time scales -- beyond the equilibration of local degrees of freedom on a scale $1/t_d$ -- changes
drastically: the slow divergence of the entropy at very large times that is observed in the quasi-MBL phase, and
associated with a slow relaxation of the system at all length scales, appears completely suppressed in the case of
a weakly disordered system.
Furthermore, the entropy saturates to a far sub-thermal value for both $L=4$ (not shown) and $L=8$ even
for the weakest disorder potential $W=0.01$. These results are
consistent with the observation of a many-body localized regime in this limit in the entanglement entropy
of highly excited eigenstates, and is also consistent with the observation of a divergent susceptibility to spin glass ordering
in Ref.~\onlinecite{yao2014}.

Increasing the hopping in the clean chain to $t_c=0.1$ (bottom panel of Fig.~\ref{fig:te}), we find that a weak disorder potential suppresses
the saturation entropy only slightly, while a sufficiently strong potential in the disordered chain is still able to
suppress the entropy for both chains. While a finite critical disorder strength necessary to drive the system into the MBL phase
appears at odds with a quasi-MBL regime in the clean limit, our numerical observation may also be due to a
large localization length compared to the available system sizes. In either interpretation, these results
are consistent with the observation of a possible MBL phase
in a similar parameter regime in the entropy of eigenstates, see Fig.~\ref{fig:hopbox}.

\section{Conclusion}
\label{sec:concl}

We have proposed a system of spinless fermions on a ladder, where one chain of the ladder is translationally invariant
while the other experiences a disorder potential. This system can serve as a prototypical example to study many-body
localization effects in a system where in the limit of vanishing interactions, localized and delocalized degrees of freedom
coexist. Equivalently, the system allows us to study the effect of coupling a small bath to a many-body localized system.
Exploiting the large degree of tunability of the model, we have found that while in most of the phase diagram the system
tends towards delocalization, a many-body localized regime -- and thus localization of the bath through induced disorder
from the MBL system -- may appear in a regime where the bandwidth of the bath is small compared to the disorder strength.

In the entanglement of eigenstates, this localization of the entire system is heralded by a decreasing entanglement entropy
as a function of the interaction strength. Attempting to extrapolate to larger systems, we find a broad regime of parameters
where the trend of suppression of the entropy by interactions reverses for larger system sizes, and the
system tends towards delocalization as the system size is increased. By taking the hopping in the clean chain
very small and the disorder very strong, we obtain a regime where the entropy remains small for all system sizes accessible
to our numerics; however, we cannot answer in the affirmative whether we
obtain a genuine many-body localized phase of both chains, or rather a regime where the crossover to delocalization
takes place at an extraordinarily large system scale.
Turning to the time evolution of the entropy starting from random initial product states, a strong suppression of the
saturation entropy is observed in a similar parameter regime, providing further support for a many-body localized phase.

Finally, we comment on the intermediate non-ergodic yet delocalized phase observed in Ref.~\onlinecite{li2015} in a window of many-body
energy densities between a localized regime at low energies, and a fully ergodic delocalized regime at high energies. Since
the crossover regime in the problem considered here is parametrically very large, reliably observing such a regime appears
very challenging and has not been attempted systematically. However, we point out that tuning into such a regime as a
function of energy is less natural in the model considered here, since in the single-particle limit localized and delocalized
states coexist at all energies, whereas in the model of Ref.~\onlinecite{li2015} the energy density can be understood as tuning
the relative number of localized and delocalized orbitals in the non-interacting limit. A more natural parameter to replicate
the phase diagram in our model would thus be the relative filling of the clean and disordered chains. However, given the
small range of system sizes available to us, changing the average inter-particle distance is likely to incur larger finite-size
corrections.

\acknowledgements

We thank Chris Laumann, Sankar Das Sarma, Sarang Gopalakrishnan and Sid Parameswaran for insightful discussions.
This research was supported in part by the National Science Foundation, under Grant No.\ DMR-14-04230 (JRG), and by the Caltech Institute of Quantum Information and Matter, an NSF Physics Frontiers Center with support of the Gordon and Betty Moore Foundation (JRG).

This material is based upon work supported by the National Science Foundation Graduate Research Fellowship under Grant No. DGE 1144085. Any opinion, findings, and conclusions or recommendations expressed in this material are those of the authors(s) and do not necessarily reflect the views of the National Science Foundation.

We acknowledge support from the Center for Scientific Computing at the CNSI and MRL: an NSF MRSEC (DMR-1121053) and NSF CNS-0960316.
ACP was supported by the Gordon and Betty Moore Foundation's EPiQS Initiative through Grant GBMF4307.

\bibliography{mbleth}

\end{document}